\documentclass[amsmath,twocolumn,amssymb,showpacs,pra]{revtex4}
\usepackage{graphicx}
\usepackage{psfrag} 
\usepackage{epsfig}
\usepackage{dcolumn}
\usepackage{rotating}

\input{epsf}
\usepackage{mathrsfs}
\usepackage{theorem}

\usepackage{bm}
\usepackage{multirow}

\newcommand{\ket}[1]{\left| #1 \right\rangle}

\newcommand{\braket}[2]{\langle #1 | #2\rangle}

\newcommand{\beq}{\begin{equation}}
\newcommand{\eeq}{\end{equation}}
\newcommand{\beqy}{\begin{eqnarray}}
\newcommand{\eneqy}{\end{eqnarray}}
\newcommand{\bal}{\begin{align}}
\newcommand{\eal}{\end{align}}
\begin{document}

\title{Optimised control of Stark-shift-chirped rapid-adiabatic-passage in a
$\Lambda$-type three-level 
system}
\author {Johann-Heinrich Sch\"onfeldt}
\altaffiliation{E-mail: {\tt jhschonf@ics.mq.edu.au}}
\author{Jason Twamley}
\author {Stojan Rebi\'c}
\affiliation{Centre for Quantum Computer Technology, Physics Department,
Macquarie University, 
Sydney, NSW 2109, Australia}

\begin{abstract}
Inhomogeneous broadening of energy levels is one of the principal limiting
factors for achieving ``slow'' or ``stationary'' light in solid state media by
means of electromagnetically induced transparency (EIT), a quantum version of
stimulated Raman adiabatic passage (STIRAP). Stark-shift-chirped
rapid-adiabatic-passage (SCRAP) has been shown to be far less sensitive to
inhomogeneous broadening than STIRAP, a population transfer technique to which
it is closely related. We further optimise the pulses used in SCRAP to be even
less sensitive to inhomogeneous broadening in a $\Lambda$-type three-level
system. The optimised pulses perform at a higher fidelity than the standard
gaussian pulses for a wide range of detunings (i.e. large inhomogeneous
broadening).
\end{abstract}

\pacs{32.80.Qk, 33.20.Bx, 33.80.Be, 42.50.Hz, 42.50.Gy}
\maketitle
\indent 

\section{Introduction}\label{intro}
Atomic vapours have been extensively investigated for use as quantum information
storage media by ``slowing down'', or even ``stopping'', a light pulse carrying
quantum information. This is achieved by mapping the quantum state of the light
to a long-lived spin state in an ensemble of atoms by means of a reversible
process: electromagnetically induced transparency (EIT) \cite
{harris97,hau99,phillips01,fleischhauer02,lukin03}, which is closely related to
an adiabatic passage technique known as Stimulated Raman Adiabatic Passage
(STIRAP). The homogeneity of the atomic vapour atoms means that the light pulses
used in EIT experiments in these systems can be tuned to specific transitions in
the atoms and that these light fields will then interact strongly with all the
atoms in the ensemble. 

Solid-state implementations of stored light by means of EIT may have a number of
potential advantages over atomic vapour implementations. They have potential to
greatly reduce (possibly eliminate) limitations on the storage lifetime, which
is mainly due to atomic diffusion and Doppler velocities in atomic vapour
systems. A higher atomic density has the potential to yield a stronger
interaction between the light and atomic ensemble, whilst  being more compact
and simple to use/manufacture may indicate great scalability \cite{wu08} on the
part of solid-state implementations. EIT and slow light have been 
demonstrated in rare-earth doped semiconductors e.g. $Pr$ doped $Y_2SiO_5$
\cite{turukhin01, longdell05}, whereas only EIT has thus far been shown in
nitrogen-vacancy colour centers in diamond \cite{wei99, hemmer01}. Solid-state
media are, however, not without their own problems, the biggest of which is
inhomogeneous broadening of the energy levels, which leads to a reduction in the
number of atoms/centers that will be resonant with the incident optical field.
This broadening occurs because each atom/center experiences a different
electronic environment due to a number of factors e.g. the local strains in the
crystal lattice. 

Recently a variation of STIRAP (the basis of EIT), namely Stark-shift-chirped
rapid adiabatic passage 
(SCRAP), has been shown to perform coherent population transfer with a high
fidelity for a range of 
different detunings in a $\Lambda$-type three-level system \cite{rangelov05}.
That is, SCRAP is far more 
robust when there is large inhomogeneous broadenings present in the ensemble. In
SCRAP, a separate 
(pulsed) field is added to the STIRAP fields to induce Stark shifts in the
energy levels and thus bring the 
required, initially off-resonant, transitions on resonance at specific times. We
propose that a quantum 
version of SCRAP could surmount some of the limitations that inhomogeneous
broadening places on 
``slow'' light in solid state systems. We will examine this further in future
work.

In this work we make use of optimum control techniques similar to Khaneja
\emph{et a.l} (2005) \cite
{khaneja05}, in order to optimise the standard SCRAP pulses so as to minimise
the decrease in fidelity 
brought on by inhomogeneous broadenings of the transitions. To measure this we
simulate the SCRAP 
process population transfer  between two long lived ground states which
experience large detunings due 
to inhomogeneous broadening. Our main result is that we can improve the average
fidelity of population 
transfer over a wide range of detunings for both the ground to excited state
detuning and the ground to 
target state detuning (two-photon detuning).  The optimal control pulses are
thus tailored to provide 
effective state transfer (and thus EIT), in the presence of large inhomogeneous
broadening.

The paper is arranged as follows: In section \ref{scrapsection} the SCRAP
technique is introduced. 
Section \ref{optimise} describes the optimisation methods used. Section
\ref{results} shows the results 
for the optimised pulses. Conclusions are drawn in Section \ref{conclusions}.

\section{SCRAP}\label{scrapsection}

\begin{figure}[tbp]  %Figure 1 %
\begin{center}
\setlength{\unitlength}{1cm}
\begin{picture}(7,7)
\put(-0.8,0.5){\includegraphics[width=8.4\unitlength]{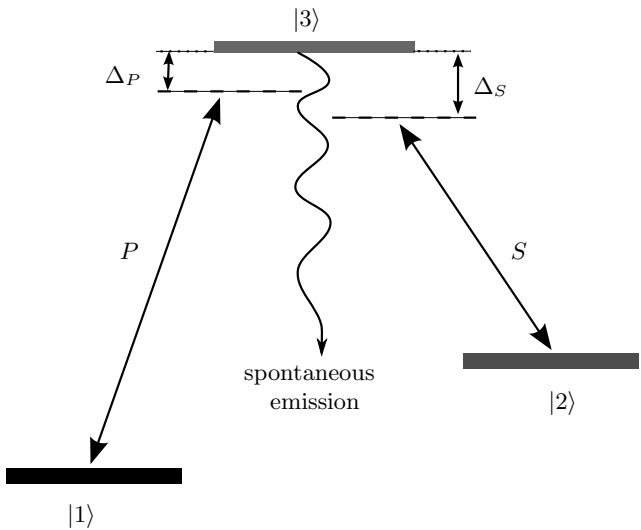}}
\put(-0,-0){$\ket{1}$}
\put(6.4,1.5){$\ket{2}$}
\put(3,6.6){$\ket{3}$}
\put(2.35,1.9){spontaneous}
\put(2.7,1.5){emission}
\put(0.5,5.85){$\Delta_P$}
\put(5.4,5.7){$\Delta_S$}
\put(0.7,3.5){$P$}
\put(5.9,3.5){$S$}
\end{picture}
\end{center}
\caption{ The $\Lambda$-type three-level energy scheme. States $|1\rangle$ and
$|2\rangle$ are 
coupled by the pump pulse $P$ which has a detuning $\Delta_P$ from being exactly
on resonance. 
Similarly states $|2\rangle$ and $|3\rangle$ are coupled by the Stokes pulse $S$
which has a detuning $
\Delta_S$. State $|2\rangle$ is short lived with spontaneous emission occurring 
out of the system.}
\label{lambda}
\end{figure}

The Stark-chirped rapid adiabatic passage technique (SCRAP) was first proposed
by Yatsenko {\em et 
al.} \cite{yatsenko99}, and implemented by Rickes {\em et al.} \cite{rickes00},
in two-level systems as an 
efficient method for complete population transfer between two states. It was
later shown by Rangelov 
\emph{et al.} (2005) \cite{rangelov05}, that a Stark shifting pulse can also be
used to achieve complete 
population transfer through adiabatic passage in a three-level system, thus
providing an alternative to 
STIRAP. 

The $\Lambda$-type three-level system (Figure \ref{lambda}), is comprised of two
long lived ground 
states, one the initially populated state ($|1\rangle$), and the other the
target state ($|3\rangle$), and an 
excited state ($|2\rangle$). In general the excited state has a short life time:
STIRAP is effective at 
complete population transfer since it avoids populating the excited state by
having the system evolve 
along a dark-state. In both SCRAP and STIRAP states $|1\rangle$ and $|2\rangle$
are coupled by the 
``pump'' laser pulse whilst states $|3\rangle$ and $|2\rangle$ are coupled by
the ``Stokes'' laser pulse. 
The frequencies of these two classical laser fields are typically not exactly on
resonance with their 
respective transitions, and have detunings $\Delta_P$ and $\Delta_S$ for the
pump and Stokes lasers 
respectively. Herein lies the advantage that SCRAP has over STIRAP: STIRAP
requires exact two-photon 
resonance ($\Delta_P-\Delta_S=0$) in order to be effective, whereas SCRAP has a
larger tolerance for 
two-photon detuning. In SCRAP a third strong far-off-resonance laser pulse, the
Stark pulse, is also 
applied. The Stark pulse induces a Stark shift in the energy of the excited
state (essentially leaving the 
energy of the lower levels unchanged), thus bringing the initially off-resonant
transitions into resonance. 
The Stark shift causes the diabatic energy of state $|2\rangle$ to cross those
of states $|1\rangle$ and $|
3\rangle$, allowing population transfer from $|1\rangle$ to $|2\rangle$ and then
to $|3\rangle$, 
completing the population transfer. It is thus clear that the diabatic energies
of state $|1\rangle$ and $|2
\rangle$ must cross before the crossing of energies of states $|2\rangle$ and
$|3\rangle$. It is also clear 
that decay out of the excited state will play a role in SCRAP, as opposed to
STIRAP.

In the rotating wave approximation, the Hamiltonian of the $\Lambda$-type
three-state system depicted 
in Figure \ref{lambda} is
\begin{align}\label{starkhamil}
H(t) = \frac{\hbar}{2} 
\begin{bmatrix}
0 & \Omega_P(t) & 0 \\
\Omega_P(t) & 2\left(\Delta_P + S_2\left(t\right)\right) - i\Gamma & \Omega_S(t)
\\
0 & \Omega_S(t) & 2\left( \Delta_P-\Delta_S\right) \\
\end{bmatrix},
\end{align}
where $\Omega_P(t)$ and $\Omega_S(t)$ are respectively the pump and Stokes laser
field Rabi 
frequencies and $S_2\left(t\right)$ is the Stark shift in the energy of the
excited energy level $|2\rangle$ 
due to a third far-off-resonant laser pulse (Stark pulse). For the examples here
it is assumed that the 
Stark shift is negative, $S_2\left(t\right)< 0$. The detunings of the pump and
Stokes fields are $\Delta_P$ 
and $\Delta_S$ respectively. The imaginary term $i\Gamma$ describes the losses
from $|2\rangle$ due 
to spontaneous radiative decay out of the three-level system. The Stark shifts
of the energy levels for 
states $|1\rangle$ and $|3\rangle$ are not included here because Stark shifts in
ground and meta-stable 
states tend to be much smaller than those of excited states. 

In Rangelov \emph{et al.} (2005) \cite{rangelov05}), gaussian pulse shapes were
used for all the pulses, 
with identical peak values, $\Omega_0$, for the Rabi frequencies of the pump and
Stokes pulses,
\beq\label{pumppulse}
\Omega_p\left(t\right) = \Omega_0 e^{-\left(t-\tau_p\right)^2/T_P^2},
\eeq
\beq\label{stokespulse}
\Omega_s\left(t\right) = \Omega_0 e^{-\left(t-\tau_s\right)^2/T_S^2},
\eeq
\beq\label{starkpulse}
-S_2\left(t\right) = S\left(t\right) = S_0 e^{-t^2/T_{St}^2}.
\eeq
The peak of the Stark pulse (maximum Stark shift of $S_0$) is taken to be at
$t=0$, and as such the 
pump and Stokes pulses peak at times $\tau_p$ and $\tau_s$ respectively. The
pulse durations are 
determined by $T_P$, $T_S$ and $T_{St}$, where it was taken that the pump and
Stokes pulses have 
equal duration $T_P=T_S$ and the Stark pulse has twice their duration
$T_{St}=2T_P$. The unit of time 
was defined as $T_P$ and the unit of frequency as $1/T_P$. These gaussian pulses
served as 
exemplars for the initial pulses used in our optimising routine, see section
\ref{optimise}. 

The diabatic energies of the states are
\beq\label{diabaticenergies}
E_{|1\rangle} = 0,
\eeq
\beq\label{state2energy}
E_{|2\rangle} = 2\left(\Delta_P + S_2\left(t\right)\right),
\eeq
\beq\label{state3energy}
E_{|3\rangle} = 2\left( \Delta_P-\Delta_S\right),
\eeq
which with the condition that the Stark shift is negative ($S_2\left(t\right)<
0$), dictates that the diabatic 
energy of state $|2\rangle$, $E_{|2\rangle}$, will only cross those of states
$|1\rangle$ and $|3\rangle$ 
when
\beq\label{levelxcondit1}
S_0>\Delta_P>0,
\eeq
and
\beq\label{levelxcondit2}
S_0>\Delta_S>0.
\eeq
With the condition that the Stark shift is negative two distinct situations can
arise: the two-photon 
detuning can be negative
\beq\label{case1}
\Delta_P>0>\left(\Delta_P-\Delta_S\right)>\Delta_P-S_0,
\eeq
or positive
\beq\label{case2}
\Delta_P>\left(\Delta_P-\Delta_S\right)>0>\Delta_P-S_0.
\eeq
In the examples used here to explain the SCRAP technique, and for the pulse
optimisation, only the first 
case \eqref{case1}, where the two-photon detuning is negative will be presented
($\Delta_P<\Delta_S$). 
In the case that the two-photon detuning is positive \eqref{case2}, the order of
the pulses in standard 
SCRAP must be run in reverse to what will be shown here, \cite{rangelov05}.  

\begin{figure}[htbp]  %Figure 2 %
\begin{center}
\setlength{\unitlength}{1cm}
\begin{picture}(8,6.5)
\put(0.1,0.5){\includegraphics[height=5.8\unitlength]{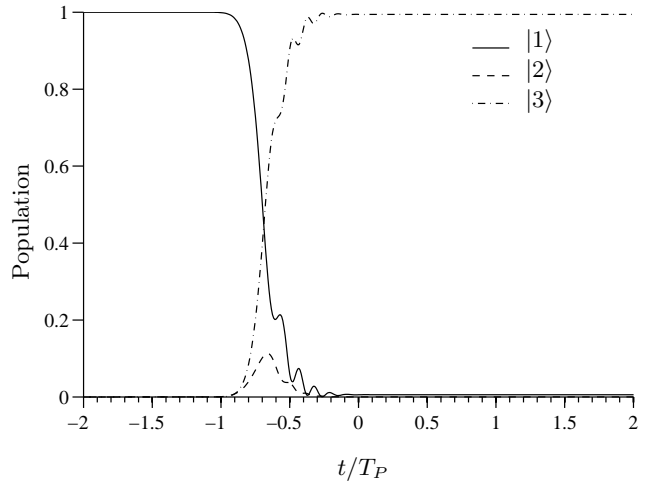}}
\put(4,-0.1){$t/T_P$}
\put(-0.1,2.5){\begin{rotate}{90}Population\end{rotate}}
\put(6.5,5.6){$\ket{1}$}
\put(6.5,5.2){$\ket{2}$}
\put(6.5,4.8){$\ket{3}$}
\end{picture}
\end{center}
\caption{The evolution of the population in the 3-state-system (state $\ket{1}$
solid line, $\ket{2}$ dashed 
line and $\ket{3}$ dash-dotted line) where the pulses used were the original
gaussians with the following 
parameters: $\Delta_P=30/T_P$, $\Delta_S=45/T_P$, $S_0=200/T_P$,
$\Omega_0=50/T_P$, 
$T_S=T_P$, $T_{St}=2T_P$, $\tau_p=-T_P$, $\tau_s=-2T_P$. There was no decay out
of state $\ket{2}
$, ($\Gamma=0$).}
\label{population}
\end{figure}

Our goal is to achieve efficient state transfer for as wide a variety of
detunings as possible, for which 
SCRAP is the ideal technique. The population transfer for the ideal situation
with no detuning, no decay, 
and using pulse parameters out of \cite{rangelov05} is shown in Figure
\ref{population}. In such a 
situation with no detuning STIRAP would be the preferred technique. For a more
detailed explanation of 
three state SCRAP please refer to section III in Rangelov \emph{et al.} (2005)
\cite{rangelov05}.

\section{Optimisation through optimal control}\label{optimise}

The above mentioned SCRAP process has the advantage that it can tolerate large
two-photon 
detunings, but unlike STIRAP it suffers from decay from state $|2\rangle$. In
this section we will use an 
optimisation technique, based on the GRAPE  algorithm by Khaneja \emph{et al.}
(2005) \cite
{khaneja05}, to optimise the transfer fidelity for a set of detunings. The state
of the three level system is 
characterised by the density operator $\rho\left(t\right)$ with the
Liouville-von Neuman equation of 
motion
\beq\label{LvN}
\dot{\rho}\left(t\right) = -i\left[\left(H_0 + \sum^m_{k=1} u_k\left(t\right)H_k
\right), \rho\left(t\right)\right].
\eeq
Where $H_0$ is the free evolution Hamiltonian and the $H_k$ are the Hamiltonians
corresponding to the 
$m$ control fields (the Stokes, pump and Stark fields). $u\left(t\right) =
\left(u_1\left(t\right), u_2\left(t
\right),  \dots, u_m\left(t\right) \right)$ is the vector of control amplitudes.
We discretised the transfer time 
$T$ into $N$ steps of length $\Delta t = T/N$, and assume that the amplitude for
each control field is 
constant during each time step. Instead of optimising these amplitudes directly,
as described in \cite
{khaneja05}, we define each amplitude vector in terms of $q$ gaussians,
\beq\label{yous}
u_k\left(j\right) = \sum^q_{n=1} h_{n,k} \exp\left[-\left(j\Delta t
-\tau_{n,k}\right)^2/\sigma^2_{n,k}\right],
\eeq
that sum to create the pulse for the specific control field, and optimise the
parameters of these 
gaussians. The aim is to find the parameters ($h_{n,k}, \tau_{n,k},
\sigma_{n,k}$) that will, given the 
initial density operator $\rho\left(0\right) = \rho_0$, maximise the overlap of
the density operator after a 
time $T$, $\rho\left(T\right)$, with a target density operator $C$. The overlap
is measured by the 
standard inner product, thus the performance index $\Phi_0$ is given by
\beq\label{phi0}
\Phi_0 = \braket{C}{\rho\left(t\right)}.
\eeq
During each time step $j$ the evolution of the system is given by the propagator
\beq\label{propagator}
U_j = \exp\left[ -i\Delta t\left( H_0 + \sum^m_{k=1} u_k\left(j\right)H_k
\right)\right].
\eeq
The performance index can then be written as
\beq\label{goal}
\Phi_0 = \braket{\underbrace{U^\dagger_{j+1}\dots U^\dagger_N C U_N \dots
U_{j+1}}_{\lambda_j}}
{\underbrace{U_j\dots U_1\rho_0 U^\dagger_1\dots U^\dagger_j}_{\rho_j}}.
\eeq
From \cite{khaneja05} we have that 
\beq\label{goa1l}
\frac{\delta\Phi_0}{\delta u_k\left(j\right)} = -\braket{\lambda_j}{i\Delta t
\left[H_k,\rho_j\right]},
\eeq
but we are interested in the gradient with regard to $h_{n,k}$, $\tau_{n,k}$ and
$\sigma_{n,k}$:
\bal\label{}
\frac{\delta\Phi_0}{\delta h_{n,k}} = \sum^N_j & \left[\frac{\delta\Phi_0}{\delta
u_k\left(j\right)}\times\frac
{\delta u_k\left(j\right)}{\delta h_{n,k}}\right] \notag\\
= \sum^N_j -&\braket{\lambda_j}{i\Delta t \left[H_k,\rho_j\right]}\times \notag\\
& \exp\left[-\left(j\Delta t -\tau_{n,k}\right)^2/\sigma^2_{n,k}\right],
\end{align}
and similarly
\bal\label{}
\frac{\delta\Phi_0}{\delta \tau_{n,k}} = \sum^N_j -&\braket{\lambda_j}{i\Delta t
\left[H_k,\rho_j\right]}\times 
\frac{2\left(j\Delta t -\tau_{n,k}\right)}{\sigma^2_{n,k}}\times\notag\\
& h_{n,k}\exp\left[-\left(j\Delta t -\tau_{n,k}\right)^2/\sigma^2_{n,k}\right],
\\
\frac{\delta\Phi_0}{\delta \sigma_{n,k}} = \sum^N_j -&\braket{\lambda_j}{i\Delta
t \left[H_k,\rho_j\right]}
\times \frac{2\left(j\Delta t
-\tau_{n,k}\right)^2}{\sigma^3_{n,k}}\times\notag\\
& h_{n,k}\exp\left[-\left(j\Delta t -\tau_{n,k}\right)^2/\sigma^2_{n,k}\right].
\end{align}
The performance $\Phi_0$ increases if we choose
\bal\label{asd}
h_{n,k} = h_{n,k} + \epsilon\frac{\delta\Phi_0}{\delta h_{n,k}},
\end{align}
with $\epsilon$ a small step size, and similarly for $\tau_{n,k}$ and $
\sigma_{n,k}$. 

We made use of a Matlab routine "minFuncBC'', written by Mark Schmidt
(\url{http://www.cs.ubc.ca/
~schmidtm/Software/minFunc.html}), to perform the final optimisation step
\eqref{asd}. It makes use of a 
quasi-Newton method, the BFGS method, and accepted as input $\Phi_0$ and the
vector of derivatives 
with respect to the optimising parameters, $\left[ \frac{\delta\Phi_0}{\delta
h_{n,k}}, \frac{\delta\Phi_0}
{\delta \tau_{n,k}}, \frac{\delta\Phi_0}{\delta \sigma_{n,k}}\right]$.

\section{Results}\label{results}

In order to optimise the pulses for as large a detuning space as possible, an
initial single point in the 
detuning space was chosen by searching for the detuning point with optimised
pulses that performed the 
best over the whole chosen detuning space. This entailed calculating the
fidelity, using equation \eqref
{phi0}, for each point in the detuning space we would like to optimise, and then
taking the average. Once 
the first point was found a second point would be chosen by searching for the
second point that, together 
with the first, would result in the best optimised pulses. This process would be
repeated until no 
additional points result in better pulses. The total performance function used
for evaluating the efficiency 
of the pulses during the optimisation on a number $d$ of detunings was taken as
the average of the 
performance functions \eqref{phi0} for each of the $d$ detuning points,
\bal\label{}
\Phi = \frac{1}{d}\sum_x^d \Phi_0\left(\Delta_P^x,\Delta_S^x\right).
\end{align}

\begin{figure}[tbp]  %Figure 3 %
\begin{center}
\setlength{\unitlength}{1cm}
\begin{picture}(8,6.3)
\put(0.5,0.5){\includegraphics[height=5.5\unitlength]{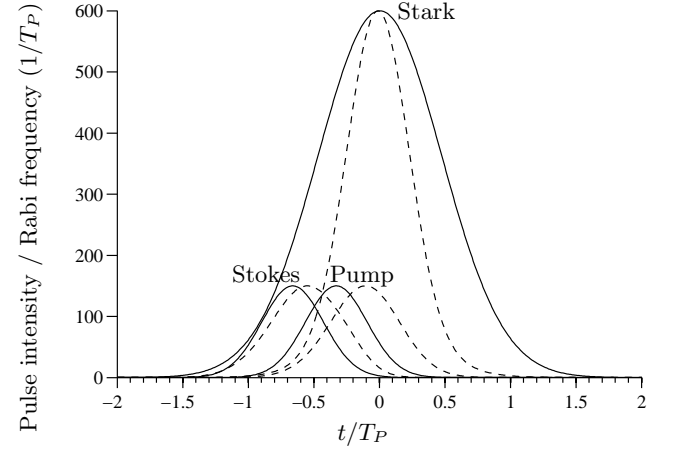}}
\put(4,0.1){$t/T_P$}
\put(-0,0.2){\begin{rotate}{90}Pulse intensity / Rabi frequency
($1/T_P$)\end{rotate}}
\put(4.8,5.7){Stark}
\put(3.9,2.2){Pump}
\put(2.6,2.2){Stokes}
\end{picture}
\end{center}
\caption{The original gaussian pulses (solid lines) with $S_0=200/T_P$,
$\Omega_0=50/T_P$, 
$T_S=T_P$, $T_{St}=2T_P$, $\tau_p=-T_P$, $\tau_s=-2T_P$. The dashed lines are
the pulses 
optimised for the detunings shown in Figure \ref{optimised}.}
\label{gaussians}
\end{figure}

For the initial pulses used in our optimal control routine we used 9 gaussians
of equal amplitude to 
approximate each of the original pulse shapes \eqref{pumppulse},
\eqref{stokespulse} and \eqref
{starkpulse}. That is, in equation \eqref{yous}, $q=9$ and $k \in\{P,S,St\}$
(the probe, Stokes and Stark 
fields). The parameters of these gaussians are given in Table \ref{table}.
\begin{table}[tbp]
	\begin{ruledtabular}
	\begin{tabular}{r|*{5}{c}}
	$n$ & 1 & 2 & 3 & 4 & 5\\ \hline
	$\tau_{n,k}$ & $\tau_k$ & $\tau_k-0.15T_k$ & $\tau_k+0.15T_k$ &
$\tau_k-0.4T_k$ & $\tau_k
+0.4T_k$  \\
	$\sigma_{n,k}$ & $\sqrt{0.2}T_k$ & $\sqrt{0.2}T_k$ & $\sqrt{0.2}T_k$ &
$\sqrt{0.25}T_k$ & $\sqrt
{0.25}T_k$ \\ \hline
	 $n$ & & 6 & 7 & 8 & 9  \\ \hline
	 $\tau_{n,k}$ & & $\tau_k-0.55T_k$ & $\tau_k+0.55T_k$ & $\tau_k-T_k$ &
$\tau_k+T_k$  \\
	 $\sigma_{n,k}$ & & $\sqrt{0.2}T_k$ & $\sqrt{0.2}T_k$ & $\sqrt{0.32}T_k$ &
$\sqrt{0.32}T_k$  \\
	\end{tabular} 
	\end{ruledtabular}
	\caption{The parameters for each of the 9 gaussians that constitute each
of the pulses. The 
amplitudes of the gaussians for a given pulse were all eaqual:
$h_{n,P}=h_{n,S}=0.23\Omega_0$ and 
$h_{n,St}=0.23S_0$.}
	\label{table}
\end{table}
The original pulses (solid lines in Figure \ref{gaussians}) had the following
parameters: $S_0 = 200/T_P
$, $\Omega_0 = 50/T_P$, $T_S = T_P$, $T_{St }= 2T_P$, $\tau_P = -T_P$, $\tau_S =
-2T_P$, $\tau_
{St}=0$. 

The maximum of the optimised pulses was constrained to that of the original
pulses ($S_0=200/T_P$, $
\Omega_0=50/T_P$). If this is not done pulses can be made arbitrarily efficient
by simply increasing the 
maximum Rabi frequency. This is not physically possible due to adiabaticity
constraints. Furthermore the 
optimisation was constrained to prevent the optimised pulses from becoming too
narrow by setting the 
lower bound on the width of the pulses to 50\% of the original pulse width. The
optimised pulses are 
shown as the dashed lines in Figure \ref{gaussians}.

\begin{figure}[htbp]  %Figure 5 %
\begin{center}
\setlength{\unitlength}{1cm}
\begin{picture}(8.,5.5)
\put(-0.,0.2){\includegraphics[height=5.5\unitlength]{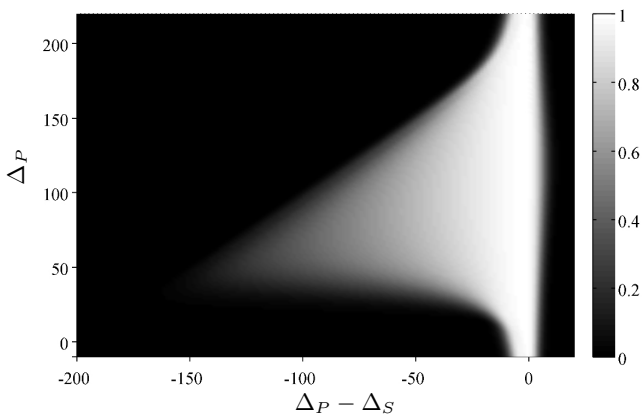}}
\put(3.5,0){$\Delta_P-\Delta_S$}
\put(-0.1,3){\begin{rotate}{90}$\Delta_P$\end{rotate}}
\end{picture}
\end{center}
\caption{The fidelity for a range of detunings, using the original gaussian
pulses for SCRAP (shown as 
the solid lines in Figure \ref{gaussians}), with $S_0=200/T_P$,
$\Omega_0=50/T_P$, $T_S=T_P$, $T_
{St}=2T_P$, $\tau_p=-T_P$, $\tau_s=-2T_P$. The decay out of state $|2\rangle$
was $\Gamma=1/T_p
$.}
\label{original}
\end{figure}

\begin{figure}[htbp]  %Figure 6 %
\begin{center}
\setlength{\unitlength}{1cm}
\begin{picture}(8,5.5)
\put(0,0.2){\includegraphics[height=5.5cm]{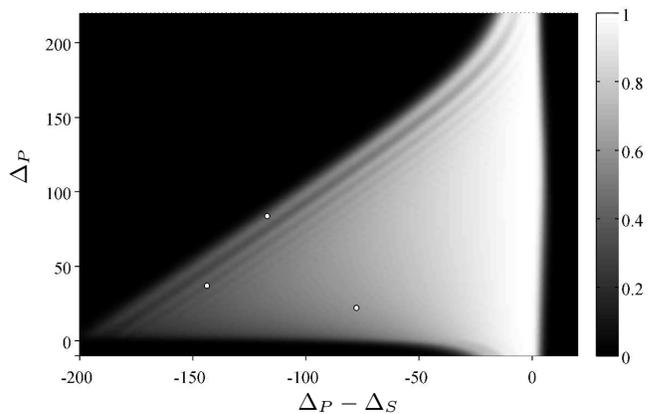}}
\put(3.5,0){$\Delta_P-\Delta_S$}
\put(-.1,3){\begin{rotate}{90}$\Delta_P$\end{rotate}}
\end{picture}
\end{center}
\caption{The fidelity for a range of detunings, using SCRAP pulses optimised for
the detunings indicated 
(white cirlcles). Optimised pulses shown as the dashed lines in Figure
\ref{gaussians}. The decay out of 
state $|2\rangle$ was $\Gamma=1/T_p$.}
\label{optimised}
\end{figure}

\begin{figure}[htbp]  %Figure 7 %
\begin{center}
\setlength{\unitlength}{1cm}
\begin{picture}(8,5.5)
\put(0,0.2){\includegraphics[height=5.5cm]{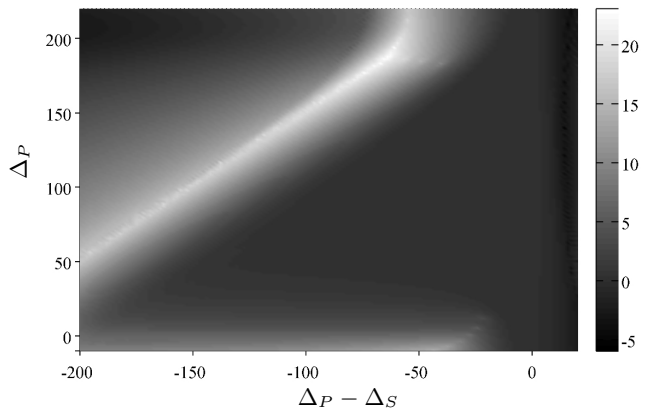}}
\put(3.5,0){$\Delta_P-\Delta_S$}
\put(-.1,3){\begin{rotate}{90}$\Delta_P$\end{rotate}}
\end{picture}
\end{center}
\caption{The $\log_{10}$ of the percentage increase in fidelity between standard
SCRAP (Figure \ref
{original}) and optimised SCRAP (Figure \ref{optimised}) for each point in the
detuning space.}
\label{logincrease}
\end{figure}

The optimised pulses were then used to evaluate the fidelity of the population
transfer from state $|1
\rangle$ to state $|3\rangle$ for a range of detunings, keeping in mind that the
chosen pulse ordering 
restricts the choice of detunings \eqref{case1}. As can be seen from Figures
\ref{original} and \ref
{optimised} the total ``area'' of detunings where population transfer is at all
possible (the parts that are 
not black) is greatly increased when using the optimised pulses. The efficiency
of the pulses can be 
gauged in a number of ways: 
\begin{enumerate}
\item measuring the normalised ``area'' in detuning space where the pulses
result in a fidelity greater 
than 0.8 for population transfer. For the original SCRAP pulses $A^{ori}_{>0.8}
= 0.131$, whilst for the 
optimised pulses $A^{opt}_{>0.8} = 0.178$ was obtained, an increase of 35.5\%.
\item measuring the average fidelity over the whole detuning space, i.e. the sum
of the fidelities for each 
point, divided by the number of points. For the original SCRAP pulses
$F^{ori}_{av}=0.219$, whilst for 
the optimised pulses $F^{opt}_{av}=0.321$, an increase of 46.6\%.
\item in Figure \ref{logincrease} a plot of the percentage increase for each
point in the detuning space is 
presented. The $\log_{10}$ of the percentage increase is used since some points
had an initial fidelity of 
virtually zero and ended with a percentage increase of $\approx 10^{23}\%$.
\end{enumerate}

\section{Conclusions}\label{conclusions}
It is clear that Stark-shift Chirped Rapid Adiabatic Passage (SCRAP) is a useful
process to employ when 
trying to overcome inhomogeneous broadening of energy levels in a system
undergoing state transfer. 
We have shown that the standard SCRAP pulses can be optimised so that a larger
inhomogeneous 
broadening can be compensated for and so that the overall fidelity of state
transfer for a range of 
detunings will be increased. Future work will aim to translate SCRAP into the
quantum domain where the 
pump pulse is replaced by a quantum probe field carrying quantum information
that is to be stored in the 
spin coherence of the atomic system. 

\section*{Acknowledgments}
Johann-Heinrich Sch\"onfeldt is supported by an MQRES scholarship from Macquarie
University.
\newpage
\bibliography{optimalscrap}

\end{document}